\documentstyle[prd,aps,floats]{revtex}
\advance\voffset by 0.5in
\begin{document}
\draft
\newcommand{\bce}{\begin{center}} \newcommand{\ece}{\end{center}}
\newcommand{\beq}{\begin{equation}} \newcommand{\eeq}{\end{equation}}
\newcommand{\beqy}{\begin{eqnarray}}
\newcommand{\eeqy}{\end{eqnarray}} \input epsf
\renewcommand{\topfraction}{0.8}
\twocolumn[\hsize\textwidth\columnwidth\hsize\csname
@twocolumnfalse\endcsname


\title{Ultra-high energy cosmic rays, cosmological constant, 
and $\theta$-vacua}
\author{Prashanth Jaikumar and Anupam Mazumdar} 
\address{ Physics Department, McGill University, Montr\'eal, 3600,
Qu\'ebec, Canada H3A 2T8.}
\maketitle

\begin{abstract}
We propose that the origin of ultra-high energy cosmic rays beyond the 
GZK cutoff and the origin of small cosmological constant can both be 
explained by vacuum tunneling effects in a theory with degenerate vacua 
and fermionic doublets. By considering the possibility of tunneling from 
a particular winding number state, accompanied by violation of some 
global quantum number of fermions, the small value of the vacuum dark 
energy and the production of high energy cosmic rays are shown to be related.  
We predict that the energy of such cosmic rays should be at least 
$5\times 10^{14}$~GeV.
\end{abstract}

\vskip2pc]


The two outstanding puzzles of modern astro-particle physics are the 
observed value of the small cosmological constant, and the origin of 
ultra-high energy cosmic rays. The latter with energies ranging from
$10^{5}-10^{22}$~eV, ever since it was observed in the first half of 
the last century, posed an open question which attracted many new ideas 
within conventional astrophysics; from the particle spectrum of the 
Standard Model (SM) to beyond the SM~\cite{kbs}. Undoubtedly such a 
vast range of energies could never be covered by a single source for the 
origin of the cosmic rays. The observed broken power law spectrum of 
cosmic rays gradually steepens as the energy increases from 
$10^{15}$~eV, known as a {\it knee}, to $10^{18}$~eV known as an {\it ankle}, 
and subsequently flattens above $10^{18}$~eV. It is usually believed that 
the first transition in the observed spectrum reassigns the origin of 
cosmic rays from galactic to extra galactic sources. However there are 
obvious constraints on a primary particle accelerated up to $10^{18}$~eV 
if they are either charged or a heavy nuclei which interacts with 
a cosmic micro wave photon background~$T\sim 3$~K, and thus they cannot 
traverse further than a few Mpc without losing energy. This is known as 
Greisen-Zatsepin-Kuzmin (GZK) effect \cite{gzk1,gzk2}.
Therefore distant astrophysical sources which might be able to generate 
such energetic particles might not be a suitable candidate for ultra high
energy cosmic rays with energies beyond $10^{18}$~eV. A simple solution 
to this impasse is to look for a candidate which is not only capable
of producing ultra high energy cosmic rays but also avoids the GZK cut-off. 

\vskip0.1cm

Such candidates could be topological defects~\cite{hill87}, or they
could come from the decays of the primordial super heavy
particles~\cite{ellis90}. In the latter scenario the mass of the
unknown $X$ particles could be ranging from $10^{12}$~GeV and
above. However the longevity of the $X$ particles (equivalent to the
age of the Universe $\sim \tau_{X}\sim 10^{10}$~years) demands an
extraordinary suppression in their interaction. Rather interesting
solutions have been put forward~\cite{kuzmin98,hamaguchi98}. In the
latter reference it was assumed that the required life time can be
obtained by imposing discrete gauge symmetries even if $X$ is an
elementary particle. Similar ideas in string theory can be found
in~\cite{ellis92}. While in~\cite{kuzmin98}, the reason for the long
life time of the $X$ particles was explained via the instanton
mediated decay, which we shall explore here in some detail.

\vskip0.1cm

An interesting connection can be made between the abundance of the 
non-luminous cold dark matter, observed as $30\%$ of the critical 
energy density of the Universe: $\rho_{c}\approx 4\times 10^{-47}~(\rm GeV)^4$, 
and the origin of the cosmic rays (above GZK cut-off) provided that 
the cold dark matter constituent is $X$ particles with a mass 
$m_{X}\sim 10^{11}-10^{15}$~GeV. Such heavy particles can be produced
abundantly to match the correct cold dark matter abundance right after 
the end of inflation~\cite{chung99}, or from the direct decay of the 
inflaton~\cite{allahverdi02}.

\vskip0.1cm

On the other hand the majority of the energy density $\sim 70 \%$ is in the
form of dark energy, whose constituent is largely unknown, but usually
believed to be the {\it cosmological constant}~\cite{perlimutter99}. The 
{\it bare} and the observed cosmological constant is a severe problem, 
especially why the observed cosmological constant is so small
$\sim 2.8 \times 10^{-47}~(\rm GeV)^4$, or in other words in Planck units
$\sim 10^{-120}~(M_{\rm p})^4$ (where we use the reduced Planck mass
$M_{\rm P}\sim 2.4\times 10^{18}$~GeV). One would naively expect that even if
the bare cosmological constant can be made to be vanishing, the quantum 
loop corrections would eventually lead to quadratic divergences 
$\sim M_{\rm p}^4$. In other words why and what keeps the cosmological 
constant so small as we see today?
Despite many attempts in a conventional big bang 
cosmological set up~\cite{many}, a convincing solution is still elusive.
Recently it was pointed out in~\cite{Yokoyama02} that, the present observable
cosmological constant can be obtained if the original vacuum can be 
associated with a non-trivial winding number. It was also assumed that 
the bare cosmological constant and the vacuum energy density vanished not 
in any one specific vacuum but in the superposed, or $\theta$-vacuum at 
some high scale $\sim 4\times 10^{16}$~GeV.

\vskip0.1cm

In this paper we address some interesting cosmological consequences of the
$\theta$-vacuum.  We propose that if the $X$ particles decay can be
explained via instanton mediation due to the transition from one
non-trivial vacuum to another, then the superposed vacua can also
be responsible for generating a non-vanishing but small cosmological
constant as we observe now.  As a result, we can relate the origin of
the cosmological constant with the ultra-high energy cosmic rays. Regarding
the $\theta$-vacuum we also generalize the description of
Ref.~\cite{Yokoyama02}, where it was strictly assumed that the 
state of the Universe is solely given by the $|n=0\rangle$ vacuum.
In what follows, we assume that the quantum state
of the Universe can be a superposition of the n-vacua of the
non-trivial $SU(N)$. As we shall see we will have a concrete prediction
for the mass of the fermion, which may act as a source for the ultra-high
energy cosmic rays. We begin by considering the origin of the $\theta$-vacuum.

\vskip0.1cm

Apart from the familiar invariance principle of gauge theories under 
small gauge transformations (those connected continuously to the 
identity), it is well-known that most grand unified theories such 
as $SU(N)$ or $SO(N)$ where $N\geq 2$ are also invariant under 
large gauge transformations. Such transformations are not continuously 
connected to the identity; rather, they generate infinitely degenerate 
perturbative vacua separated by action barriers that prevent classical 
transitions between them. Quantum mechanical tunneling can lead to a 
superposed ground state of these perturbative vacua, exponentially 
smaller in energy than any of them, since the tunneling amplitude is 
itself suppressed exponentially by the height of the action barrier. 
A practical example is the QCD vacuum, where imaginary time 
(Euclidean space) solutions of minimum action (instantons) can be 
viewed as tunneling between adjacent vacua characterized by different 
winding numbers $n$~\cite{Jackiw76}. Taking into account this quantum 
tunneling, one usually writes the true vacuum state as a weighted 
superposition of all identical $|n\rangle$ vacua
\beq
|\theta\rangle = \sum_{n=-\infty}^{\infty}e^{in\theta}|n\rangle\,,
\eeq
where for QCD, bounds from neutron electric dipole moment 
studies~\cite{Harris99} suggest that $\theta_{\rm QCD}\leq 10^{-9}$. 
However, for the purpose of illustration, let us assume that $\theta$ 
is an unconstrained parameter of some $SU(2)_X$ gauge theory, 
which is broken at some high energy scale (we will generalize our main 
results for arbitrary even $N_f$, where $N_{f}$ is the number of fermionic 
doublets). Then, there are $SU(2)_X$ instantons, which represent 
tunneling solutions between the vacua with different winding numbers. 

\vskip0.1cm

Since this true ground state is lower in energy than any particular
$n$-vacuum by only an exponentially small amount, the observed small
but finite vacuum energy density can be explained if the universe has
not yet settled down into the non-perturbative $\theta$-state, but is
still in one of the perturbative vacuum states. If one calculates the
vacuum energy density in a $\theta$-state or any $n$-state for pure
gauge theory, t'Hooft's formula~\cite{tHooft76} for the one-instanton
weight shows that the contribution of large size instantons
diverges. The vacuum energy density in any $|n\rangle$ vacua can be
expressed as~\cite{Yokoyama02} 
\beq 
\rho_v=\langle n|H|n\rangle=2Ke^{-S_0}\,, 
\eeq 
where $S_0$ is the classical instanton action $8\pi^2/g^2$ and $g$ is the
coupling constant of the $SU(2)_X$ theory. Note that we assume the
bare cosmological constant to vanish in the $\theta=0$ vacuum. 

\vskip0.1cm

In the dimensional regularization scheme~\cite{tHooft76}, 
\beqy 
K=2^{10}\pi^6~g^{-8}\int
\frac{dR}{R^5}&& \exp[-\frac{8\pi^2}{g^2(\mu_0)}+ \frac{22}{3}\ln
(\mu_0R)\\ \nonumber &+&6.998435]\,.  
\eeqy 
The integration over sizes $R$ diverges but this can be tamed by 
introducing a physical cutoff scale $\mu_0\sim M$, set by an $SU(2)$ 
scalar doublet $\Phi$ with a potential $V(\Phi)=\lambda(|\Phi|^2-M^2/2)^2$, 
a la t'Hooft~\cite{tHooft76}.  In this case, the pure gauge theory
instanton should be replaced by the constrained instanton
solution~\cite{Affleck81}, and the vacuum energy density in the
$n$-vacuum is given by
\beq 
\rho_v=\langle
n|H|n\rangle\sim\left(\frac{8\pi}{g^2}\right)^4 M^4e^{-\frac{8\pi^2}{g^2}}\,.
\label{vacenergy} 
\eeq 
If $M$ is larger than $H_{inf}$ during inflation (for chaotic type inflation
model $H_{inf} \sim 10^{13.5}$~GeV \cite{linde}) , the inflation will 
naturally tend to wipe out inhomogeneities in $\Phi$ over a Hubble volume. 
We will show that this argument constrains the size of the instanton.

\vskip0.1cm

Let us assume, for the purpose of illustration, $3$ degenerate vacua, 
$|n=0\rangle, |n=\pm 1\rangle$. Note that we allow for the Universe to be in a state other than the non-perturbative $|\theta\rangle$-state. 
The ground state of the Universe 
would then be
\beq
|G\rangle=\frac{1}{\sqrt{3}}(|0\rangle+|1\rangle+|-1\rangle)\,,\label{gstate}
\eeq
and this would have a larger energy than if all winding number states
$|n=(-\infty,\infty)\rangle$ were included (the energy of the 
$\theta$-vacuum is the least). Since the $|\theta\rangle$-state is 
also an energy eigenstate unlike the $|n\rangle$-state, we can, 
expand each of the three $|n\rangle$ states in terms of the 
$|\theta\rangle$-state. Given that 
$\langle\theta|H|\theta\rangle=\rho_v(1-{\rm cos}~\theta)$~\footnote{This expression holds in the dilute gas approximation~\cite{Callan76}, corresponding to $MR\leq 1$, which is the regime we are working in (see below).}the 
probability distribution function of the vacuum energy has a peak 
at $2\rho_v$, corresponding to $\theta=\pi$~\footnote{In QCD, 
$\theta=\pi$ is a CP-conserving value, though here the $\theta$ 
is unrelated to strong interactions.}, as shown in~\cite{Yokoyama02}. 
This peak can be interpreted as the value of the finite dark energy 
observed today. 
\vskip0.1cm

The inclusion of massive spin-1/2 fermions in this $SU(2)_X$ gauge 
theory which admits instanton solutions, has interesting physical 
consequences. t'Hooft has shown that for fermion mass $m\ll 1/R$, 
where $R$ is the typical instanton size, $N_f$ species of spin 1/2 
fermions contribute a factor~\cite{tHooft76}
\beq
A_f=(R m)^{N_f}\exp[-\frac{2}{3}N_f \ln(MR)+0.291746N_f]\,,
\eeq
to the tunneling amplitude. Using the constrained instanton solution, 
the constant $K$ now reads
\beqy
K=A_f~2^{10}\pi^6~g^{-8}\int \frac{dR}{R^5}&& \exp[-\frac{8\pi^2}
{g^2(M)}-\pi^2R^2M^2\\ \nonumber
&+&\frac{43}{6} \ln\left(\frac{MR}{\sqrt{2}}\right)+6.759189]\,.
\eeqy

\vskip0.1cm

Let us suppose now that, as in~\cite{kuzmin98}, there are $2$ fermionic 
doublets $X$ and $Y$ with different global quantum numbers, that they 
acquire this mass through the spontaneous breaking of the $SU(2)_X$ 
symmetry, and that the standard model quarks and leptons carry some 
quantum numbers under $SU(2)_X$. Mixing between $X$ and $Y$ can occur 
at the quantum level via the $SU(2)_X$ anomaly, therefore, the quantum 
transition between the adjacent vacua must be accompanied by the decay 
process~\cite{kuzmin98}
\beq
X\rightarrow Y+ {\rm SM~ quarks~ and~ leptons}\,,
\eeq 
assuming that $X$ fermions have a larger mass than $Y$ fermions. 
The large mass of the $X$ and $Y$ (Kuzmin and Rubakov~\cite{kuzmin98} 
assumed $m_X\geq 10^{13}$ GeV) particles implies that the decay 
products (quarks and leptons) are highly energetic and can lead to the 
production of highly energetic hadrons which can be the constituents 
of cosmic rays. 

\vskip0.1cm

The large mass value in our case implies the hierarchy 
of scales $R^{-1}\geq M>H_{inf}$. The first condition $MR\leq 1$ 
justifies the use of the $M\neq 0$ instanton solution, while the second 
condition $M>H_{inf}$ assures that gravitational effects on instantons are 
small during the inflationary phase and after inflation as well. Including 
the fermions, from Eq.~(\ref{vacenergy}), the vacuum energy in any 
perturbative $|n\rangle$-vacuum is
\beq
\rho_v\sim\left(\frac{8\pi}{g^2}\right)^4m^2M^2e^{-\frac{8\pi^2}{g^2}}\,,
\label{cosmic}
\eeq
where we have used the fact that $MR\sim 1$. As we have allowed for a 
small probability of transition $\Gamma$ in the horizon volume, the 
condition $\Gamma H_0^{-4}\sim 1$ holds, where the tunneling rate per 
unit volume is
\beq
\Gamma\simeq\left(\frac{8\pi}{g^2}\right)^4m^2M^2e^{-\frac{16\pi^2}{g^2}}\,.
\label{gamma}
\eeq 
Using $\rho_v\sim 10^{-120}M_{\rm p}^4$ and $H_0^2=\rho_v/(3M_{\rm p}^2)$ 
(where the subscript $0$ denotes the present era), we can solve for the 
coupling
\beq
\alpha=\frac{g^2}{4\pi}\simeq \frac{1}{44.1}\,.
\eeq
Using this value of the coupling, we find from either of the 
equations Eq.~(\ref{cosmic}) or  Eq.~(\ref{gamma}) that,
\beq
m^2M^2\sim 10^{66}{\rm GeV}^4\,.
\label{msquare}
\eeq   
From the condition $\Gamma H_0^{-4}\sim 1$, we also have 
$m^2M^2\sim \alpha^4M_{\rm p}^4$. We are proposing to explain the 
ultra-high energy cosmic rays which can have energies upto the mass 
of the decaying $X$-fermion, so it follows that one can expect cosmic rays of 
energies upto ~$\sim 5\times 10^{14}$~GeV (since $M$ cannot be greater 
than $M_{\rm p}$). We also note that $M>H_{inf}$ is satisfied.
We may generalize our scenario to arbitrary number of fermion doublets 
$N_f$ (which is required to be even by the $SU(2)_X$ anomaly). 
Accordingly, Eq.~(\ref{msquare}) is modified to 
\beq
m^{N_f}M^{4-N_f}\sim 10^{66}{\rm GeV}^4\,.
\eeq
For example, for $N_f=4$, $m\approx 3\times 10^{16}$. 
For larger $N_f$, we find 
that the fermion mass scale is not considerably different. The fermion
mass is indeed the prime result of this paper which clearly shows that 
the cascade decay of the fermions can give rise to ultra-high energy cosmic
rays with energies greater than the GZK cut-off.

\vskip0.1cm

Note that in the instanton mediated decay of fermions, the predicted mass 
turns out to be heavier than $10^{13}$~GeV (for chaotic type inflation, the
inflaton mass is around $10^{13}$~GeV in order to produce the right
amplitude for the density perturbations and the spectrum \cite{linde}),
and also greater than the observed spectrum from FLY's eye \cite{fly}
and AGASA~\cite{agasa}. Current analysis seems to be suggesting
a relic fermion mass around $10^{12}$~GeV \cite{Sarkar:2001se}-$10^{14.6}$~GeV~\cite{Fodor}.
Particularly in our case, in order to excite the superheavy fermions 
at the very first instance, one has to rely on non-thermal production 
mechanism for fermions after inflation~\cite{Giudice:1999fb} (see 
also the appendix of Ref.~\cite{Bastero-Gil:2000je}). The upcoming 
experiment such as AUGER~\cite{auger} will be able to see a considerable 
number of events above the GZK cut-off which will verify or falsify the 
energy scales which we predict here. If the inflaton coupling to the
$SU(2)_{X}$ fermions is sufficiently large then the adequate abundance 
of such fermions will be the right candidate for cold dark matter.   

\vskip0.1cm

In summary, we argue that the problem of the cosmic dark energy, 
the small value of the cosmological constant, and ultra-high energy 
cosmic rays can have a common origin. We have shown that if the 
longevity of the X-particles is due to instanton-mediated decays, 
then the fermion mass, which sets the scale for the ultra-high 
energy cosmic rays, must be larger than $5\times 10^{14}$~GeV. 
Note that this is in accordance with the observed small cosmological 
constant.


\vskip1cm
The authors are thankful to Robert Brandenberger and Guy Moore for 
discussion. P. J acknowledges support from the Natural Sciences and
Engineering Research Council of Canada. A. M is a Cita-National fellow.



\begin{references}

\bibitem{kbs}
T. Kaiser, {\it Cosmic Rays and Particle Physics}, Cambridge 
University Press, (1990); P. Bhattacharjee, and G. Sigl, Phys. Rept. {\bf 327},
109 (2000).

\bibitem{gzk1}
K. Greisen, Phys. Rev. Lett. {\bf 21}, 1016 (1966).

\bibitem{gzk2}
G. T. Zatsepin, and V. A. Kuzmin, Pis'ma Zh. Eksp. Teor. Fiz. 
{\bf 4}, 114 (1966); JETP. Lett. {\bf 4}, 78 (1966).

\bibitem{hill87}
C. T. Hill, D. N. Schramm, and T. P. Walker, Phys. Rev. D {\bf 36}, 1007 
(1987); G. Sigl, D. N. Schramm, and P. Bhattacharjee, Astropart. Phys. {\bf 2},
401 (1994); V. Berezinsky, X. Martin, and A. Vilenkin, Phys. Rev. D {\bf 56},
2024 (1997); V.~Berezinsky and A.~Vilenkin, Phys.\ Rev.\ Lett.\  {\bf 79}, 
5202 (1997); V.~S.~Berezinsky and A.~Vilenkin, Phys.\ Rev.\ D {\bf 62}, 
083512 (2000).


\bibitem{ellis90}
J. Ellis, J. L. Lopez, and D. V. Nanopoulos, Phys. Lett. B {\bf 247}, 257 
(1990).

\bibitem{kuzmin98}
V. A. Kuzmin and V. A. Rubakov,
Phys. Atom. Nucl. {\bf 61}, 1028 1998; Yad. Fiz. {\bf 61}, 1122 (1998).

\bibitem{hamaguchi98}
K. Hamaguchi, Y. Nomura, and T. Yanagida, Phys. Rev. D {\bf 58}, 103503
(1998); Phys. Rev. D {\bf 59}, 063507 (1999).

\bibitem{ellis92}
J. Ellis, G. B. Gelmini, J. L. Lopez, D. V. Nanopoulos, and S. Sarkar,
Nucl. Phys. B {\bf 373}, 399 (1992); K. Benakli, J. Ellis, and 
D. V. Nanopoulos, Phys. Rev. D {\bf 59}, 047301 (1999).

\bibitem{chung99}
D. J. H. Chung, E. W. Kolb, and A. Riotto, Phys. Rev. D {\bf 59},
023501 (1999); Phys. Rev. Lett. {\bf 81}, 4048 (1998).

\bibitem{allahverdi02}
R. Allahverdi, K. Enqvist and A. Mazumdar, Phys. Rev. D {\bf 65},
103519 (2002).

\bibitem{perlimutter99}
S. Perlmutter et. al., Astrophysics. J {\bf 507}, 46 (1998); A. G. Riess
et. al., Astron. J {\bf 116}, 1009 (1998).


\bibitem{many}
A. D. Dolgov, in {\it The Very Early Universe}, ed. G. W. Gibbons, and S. T. Siklos
(Cambridge, United Kingdom, 1982);  E. Baum, Phys. Lett. B {\bf 133}, 185 
(1983); S. W. Hawking, Phys. Lett. B {\bf 134}, 403 (1984); J. B. Hartle, 
and S. W. Hawking, Phys. Rev. D {\bf 28}, 2960 (1983), A. D. Linde, Sov. 
Phys. JETP {\bf 60}, 211 (1984); A. Vilenkin, Phys. Rev. D {\bf 30}, 509 (1984);
S. Weinberg, Phys. Rev. Lett. {\bf 59}, 2607 (1987); S. Coleman, Nucl. Phys. B 
{\bf 310}, 643 (1988); J. Polchinski, Phys. Lett. B {\bf 219}, 251 (1989); 
A. Vilenkin, Phys. Rev. Lett. {\bf 74}, 846 (1995); G. Efstathiou, 
Mon. Not. R. Astron. Soc. {\bf 274}, L73 (1995); I. Zlatev, L. Wang, and 
P. J. Steinhardt, Phys. Rev. Lett. {\bf 82}, 896 (1999); P. J. Steinhardt, 
L. Wang, and I. Zlatev, Phys. Rev. D {\bf 59}, 123504 (1999). 
For a review see S.~Weinberg, Rev.\ Mod.\ Phys.\  {\bf 61}, 1 (1989).


\bibitem{Yokoyama02}
J. Yokoyama, Phys. Rev. Lett. {\bf 88}, 151302 (2002).

\bibitem{Jackiw76}
R. Jackiw and C. Rebbi, Phys. Rev. Lett. {\bf 37}, 172 (1976).

\bibitem{Harris99}
P. G. Harris {\it et al.}, Phys. Rev. Lett. {\bf 82}, 904 (1999).

\bibitem{tHooft76}
G. t'Hooft, Phys. Rev. D {\bf 14}, 3432 (1976); Erratum {\bf 18}, 2199 (1978).

\bibitem{Affleck81}
I. Affleck, Nucl. Phys. B {\bf 191}, 429 (1981); M. Nielsen and N. K. Nielsen, 
Phys. Rev. D {\bf 61}, 105020 (2000).

\bibitem{linde}
A. D. Linde, `Particle Physics And Inflationary Cosmology,
Chur, Switzerland: Harwood (1990) 362 p.

\bibitem{Callan76}
C. G. Callan, R. F. Dashen and D. J. Gross, Phys. Lett. {\bf B63}, 334 (1976).

\bibitem{Yoko02}
J. Yokoyama, Int. J. Mod. Phys. D {\bf 11}, 1603 (2002).

\bibitem{fly}
D. J. Bird et. al., Astrophys. J {\bf 511}, 739 (1999).

\bibitem{agasa}
N. Hayashida et. al., Astropart. Phys. {\bf 10}, 303 (1999).

\bibitem{Sarkar:2001se}
S.~Sarkar and R.~Toldra, Nucl.\ Phys.\ B {\bf 621}, 495 (2002).

\bibitem{Fodor}
Z. Fodor and S. D. Katz, Phys. Rev. Lett. {\bf 86}, 3224 (2001).

\bibitem{Giudice:1999fb}
G.~F.~Giudice, M.~Peloso, A.~Riotto and I.~Tkachev,
JHEP {\bf 9908}, 014 (1999).

\bibitem{Bastero-Gil:2000je}
M.~Bastero-Gil and A.~Mazumdar,
Phys.\ Rev.\ D {\bf 62}, 083510 (2000).

\bibitem{auger}
AUGER collaboration, http://auger.cnrs.fr/pages.html


\end{references}
\end{document}